\documentclass[usenatbib]{mn2e}
\usepackage{graphicx}

\title[The maximum cosmic O abundance]
      {On the maximum value of the cosmic abundance of oxygen and 
      the oxygen yield}
      
\author[L.S.Pilyugin, T.X.Thuan and J.M.V\'{\i}lchez]
       {Leonid S. Pilyugin$^{1}$, 
        Trinh X.Thuan$^{2}$
        and Jos\'{e} M. V\'{\i}lchez$^{3}$\\
     $^{1}$ Main Astronomical Observatory
            of National Academy of Sciences of Ukraine,
            27 Zabolotnogo str., 03680 Kiev, Ukraine \\
            (pilyugin@mao.kiev.ua)\\  
     $^{2}$ Astronomy Department, University of Virginia, P.O.Box 400325, 
            Charlottesville, VA 22904, USA 
            (txt@virginia.edu) \\
     $^{3}$ Instituto de Astrof\'{\i}sica de Andaluc\'{\i}a,
            CSIC, Apdo, 3004, 18080 Granada, Spain (jvm@iaa.es) \\
             }

\date{Accepted 2006 December 19. Received 2006 December 14; in original form 2006 July 27}

\begin{document}

\maketitle

\begin{abstract}
We search for the maximum oxygen abundance in spiral galaxies. Because this 
maximum value is expected to occur in the centers of the most luminous galaxies, 
we have constructed the luminosity -- central metallicity diagram for spiral 
galaxies, based on a large compilation of existing data on oxygen abundances 
of H\,{\sc ii} regions in spiral galaxies. We found that this diagram shows a 
plateau at high luminosities (-22.3 $\la$ M$_B$ $\la$ -20.3), with a constant 
maximum value of the gas-phase oxygen abundance 12+log(O/H) $\sim$ 8.87. This 
provides strong evidence that the oxygen abundance in the centers of the most 
luminous metal-rich galaxies reaches the maximum attainable value of oxygen 
abundance. Since some fraction of the oxygen (about 0.08 dex) is expected to be 
locked into dust grains, the maximum value of the true gas+dust oxygen abundance 
in spiral galaxies is 12+log(O/H) $\sim$ 8.95. This value is a factor of $\sim$ 
2 higher than the recently estimated solar value. Based on the derived maximum 
oxygen abundance in galaxies, we found the oxygen yield to be about 0.0035, 
depending on the fraction of oxygen incorporated into dust grains.  
\end{abstract}

\begin{keywords}
galaxies: abundances -- ISM: abundances -- H\,{\sc ii} regions
\end{keywords}

\section{Introduction}

It is well known that there is a large scatter in the heavy element 
content of galaxies. This is due to the fact that different galaxies 
evolve chemically at different rates. Galactic winds with different 
efficiencies can also make a significant contribution to the
scatter in metallicity in low-mass dwarf irregular galaxies. 
The oxygen abundance in the interstellar gas is usually used as a tracer 
of metallicity in late-type (spiral and irregular) galaxies. 
A question of great interest in the chemical evolution of galaxies is that of  
the maximum value of the observed oxygen abundance. Is there such a 
value? And is this maximum value equal to the 
maximum attainable value of the oxygen abundance in galaxies? 

The latter is defined by the stellar oxygen yield, i.e. the mass of oxygen 
freshly synthesized and ejected by a generation of stars relative to the mass 
locked up in low-mass stars and compact remnants. It can be derived within 
the framework of chemical evolution models of galaxies \citep{pagbook}. 
The closed-box model which neglects mass exchange between 
a galaxy and its environments gives the maximal upper bound to the 
metallicity of the gas for a given gas mass fraction \citep{edmunds90}. 
In practice, the procedure is often inverted, i.e. the measured chemical 
compositions of galaxies are used to estimate empirically the oxygen yield 
\citep[e.g.][]{garnett02,pilyuginetal04}. This is so because the theoretical 
value of the oxygen yield is not well known, due to uncertainties in 
both the oxygen production by stars of different masses and metallicities and 
the parameters of the initial mass function of stars, the relative 
birthrates of stars with different initial masses. 
Thus the uncertainty in the stellar oxygen yield prevents an 
accurate determination of the maximum attainable value of the oxygen abundance 
in galaxies through chemical evolution models of galaxies.

Can we determine that maximum value by observations? Oxygen abundances 
in the most metal-rich spiral galaxies in the samples of 
\citet{vila92,zkh,garnettetal97,vanzeeetal98}) have been estimated 
previously by 
\citet{pilyuginetal05}. Those authors found the maximum gas-phase oxygen 
abundance in the central regions of those spiral galaxies to be 12+log(O/H) 
$\sim$ 8.75, suggesting that this value may be an upper limit 
to the oxygen abundances in spiral galaxies. 

Here we carry out a search for the maximum oxygen abundance  
using a considerably larger sample of spiral galaxies. 
The strategy of our search is based on the two following considerations. 
First, \citet{lequeuxetal79} have found that, for 
irregular galaxies, the oxygen abundance correlates with the 
total galaxy mass, in the sense that the higher the total mass, 
the higher the heavy element content. Since the galaxy mass is a poorly known 
quantity, the luminosity -- metallicity (L -- Z) relation is often used 
instead of the mass -- metallicity relation.  \citet{garnettshields87} have 
indeed found that spiral disk abundances correlate 
well with galaxy luminosities. This luminosity-metallicity 
correlation has been confirmed in many studies of spiral galaxies
\citep{vila92,zkh,garnett02,pilyuginetal04,tremontietal04}. 
Second, \citet{searle71} and \citet{smith75} have 
established many years ago the presence of radial abundance gradients 
in the disks of spiral galaxies, with the maximum oxygen abundance occurring 
at their centers. Taken together, these two facts suggest that the observed 
oxygen abundance should reach its maximal value at the centers of the most 
luminous galaxies. 

In the following, we will be using these notations for the line fluxes : 
R$_2$ = $I_{[OII] \lambda 3727+ \lambda 3729} /I_{H\beta }$,
R$_3$ = $I_{[OIII] \lambda 4959+ \lambda 5007} /I_{H\beta }$, 
R = $I_{[OIII] \lambda 4363} /I_{H\beta }$, 
R$_{23}$ = R$_2$ + R$_3$. With these definitions,  
the excitation parameter P can be expressed as: 
P = R$_3$/(R$_2$+R$_3$). 
The plan of our study is as follows.  In Section 2 we determine the central 
oxygen abundance for a large sample of nearby spiral galaxies and derive
the maximum oxygen abundance. The value of the oxygen yield is estimated in 
Section 3. We summarize our conclusions in Section 4.

\section{The maximum value of the oxygen abundance}

\subsection{Abundance derivation}

The [OIII]$\lambda$4363 auroral line is not detected in the majority of 
H\,{\sc ii} regions in spiral galaxies. 
For a quarter of a century, different versions of the one-dimensional 
empirical calibration, first proposed by \citet{pageletal79}, have been used 
for abundance determination in such H\,{\sc ii} regions. 
\citet{lcal,hcal,vybor} has found that the oxygen abundances so 
determined have a systematic error, depending on the excitation parameter 
defined above.
Recently, a relationship between the observed auroral and nebular 
oxygen line fluxes, the ff relation, has been 
established \citep{ff,pilyuginetal05}. 
The ff relation allows one to estimate the flux in the [OIII]$\lambda$4363
auroral line, and hence to apply the T$_{\rm e}$ 
method to determine abundances 
in H\,{\sc ii} regions where that line 
is not detected, and where only strong oxygen line intensities are seen.  
We have also derived a new model-independent 
$t_2$ -- $t_3$ relation \citep{tt} 
\begin{equation}
t_2 = 0.72 \, t_3 + 0.26 .
\label{equation:tt}
\end{equation}
We have thus found a way to estimate the electron temperature in both in 
the O$^{++}$ and in the O$^{+}$ zones in high-metallicity H\,{\sc ii} regions 
where the [OIII]$\lambda$4363 auroral line is not detected. 
Next, we apply the method described above to derive 
new oxygen abundances in spiral galaxies. Our aim is to search for 
a maximum value of the observed oxygen abundance in galaxies.

\subsection{A plateau in the L -- Z diagram at high luminosities}

We first define our observational sample. We use the 
compilation by \citet{pilyuginetal04} of a large sample of  
published spectra of H\,{\sc ii} regions 
in nearby spiral galaxies. 
Recent measurements from \citet{bresetal04,bresstas05,crockettetal05} 
have been added to that list. 
Since the maximum value of the observed oxygen 
abundance is expected to occur at the centers 
of the most luminous galaxies, we first derive the radial distribution 
of the oxygen abundance in the disks of the 
four most luminous spiral galaxies in our 
sample: NGC~1068 with an absolute blue magnitude M$_{\rm B}$ = --22.18, 
NGC~6384 with M$_{\rm B}$ = --22.22, NGC~7331 with M$_{\rm B}$ = --22.20, 
and IC~342 with M$_{\rm B}$ = --22.27. 
The distances and luminosities of the galaxies are taken 
from \citet{pilyuginetal04}.

\begin{figure}
\resizebox{1.00\hsize}{!}{\includegraphics[angle=000]{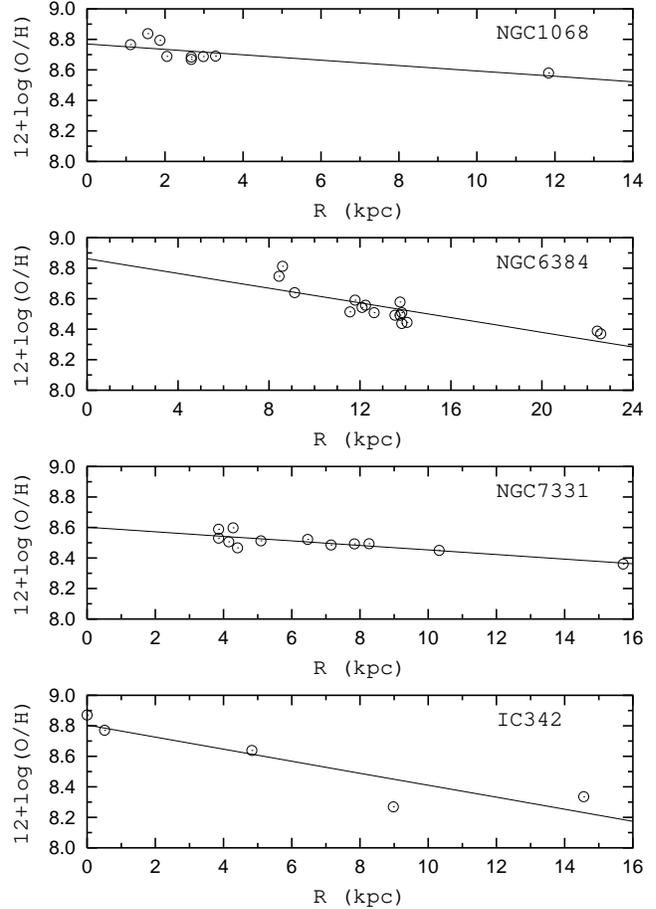}}
\caption{
The oxygen abundance as a function of galactocentric distance in the disks 
of the four most luminous spiral galaxies in our sample: NGC~1068, NGC~6384, 
NGC~7331 and IC~342. The circles are (O/H)$_{\rm ff}$ abundances in individual 
H\,{\sc ii} regions. The lines are linear least-square best fits to those 
data points.
}
\label{figure:bright}
\end{figure}

Fig.~\ref{figure:bright} shows the oxygen abundance as a function of 
galactocentric distance in the disks of the four most luminous spiral galaxies  in our sample. 
The points are (O/H)$_{\rm ff}$ abundances in individual H\,{\sc ii} regions. 
The lines are linear least-square best fits to those points.
Inspection of Fig.~\ref{figure:bright}  
shows that the central oxygen abundance in the 
most luminous spiral galaxies can be  
as large as 12+log(O/H) $\sim$ 8.87. 
Are those galaxies indeed the most oxygen-rich ones as expected 
from the L -- Z correlation? 
To clarify this point, we next derive 
the central oxygen abundances of all the spiral galaxies in our sample 
and examine their dependence on galaxian luminosity. 

\begin{figure}
\resizebox{1.00\hsize}{!}{\includegraphics[angle=000]{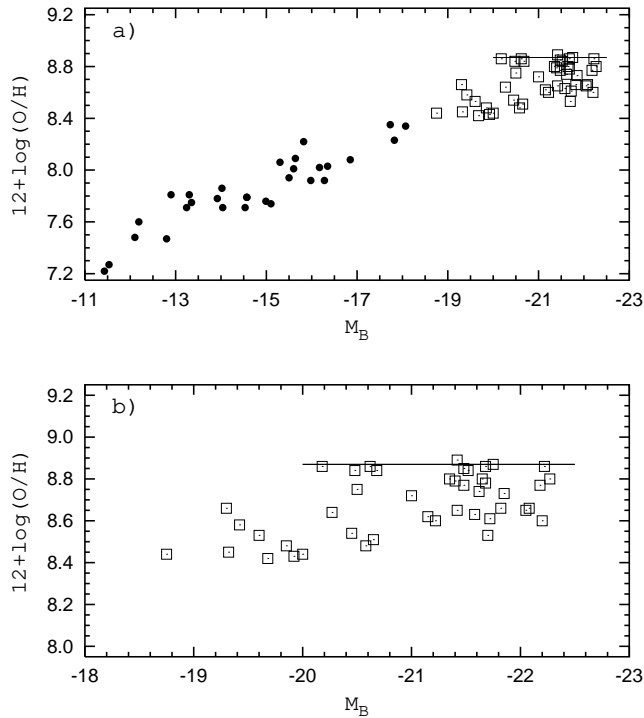}}
\caption{
The luminosity -- central metallicity diagram. 
{\it Top panel.} The open squares show the central (O/H)$_{{\rm ff}}$ 
abundances in the disks of spiral galaxies plotted against 
their luminosities. 
The filled circles denote irregular galaxies from  \citet{pilyuginetal04}.
{\it Bottom panel.} A blow-up of the high-luminosity 
part of the L -- Z diagram in the top panel. 
}
\label{figure:lz}
\end{figure}

The top panel of Fig.~\ref{figure:lz} shows  
the familiar L -- Z relation, where Z denotes here the central 
oxygen abundance.
The squares show the central (O/H)$_{{\rm ff}}$ abundances of the spiral 
galaxies in our sample. 
For the majority of galaxies, all H\,{\sc ii} regions with available
oxygen emission line measurements have been used in the abundance gradient 
analysis. For a few galaxies however, some H\,{\sc ii} regions have been 
omitted for the following reason. The $R_{\rm 23}$ -- O/H relation
is double-valued, with two distincts parts usually known as the ``lower'' and 
``upper'' branches.
The ff relation is valid only for 
H\,{\sc ii} regions which are located in 
the upper branch, with 12+log(O/H) higher 
than $\sim 8.25$. Thus, one has to know {\it a priori} 
on which of the two branches 
the H\,{\sc ii} region lies. Given that disks of spiral
galaxies show radial oxygen abundance gradients, in the sense that the oxygen
abundance is higher in the central part of the disk and decreases with
galactocentric distance, we first consider H\,{\sc ii} regions in the 
central part of disks and move outward until the radius $R^*$,
 where the oxygen 
abundance decreases to 12+log(O/H) $\sim$ 8.25. The 
exact value of $R^*$ is somewhat uncertain due to the scatter in the oxygen 
abundance values at any fixed radius. It should be emphasized that a 
misuse of ff relation in the determination
of the oxygen abundance in low-metallicity H\,{\sc ii} regions beyond $R^*$ 
may produce an artificial change of slope at $R^*$ in the 
abundance gradients \citep{bend}. 
Therefore, H\,{\sc ii} regions with galactocentric distances larger 
than $R^*$, i.e. 
those with 12+log(O/H) less than 8.25 were rejected.

The filled circles in Fig.~\ref{figure:lz} show irregular galaxies from 
\citet{lzirr01,pilyuginetal04}.  
Our L -- Z diagram for irregular galaxies is in good 
agreement with the ones derived by other authors \citep{richer95,lee03}. 
Examination of Fig.~\ref{figure:lz} shows that 
the central oxygen abundances in 
galaxies with M$_{\rm B}$ $\sim$ --20.25 are in the same range as those of 
galaxies with M$_{\rm B}$ $\sim$ --22.25, i.e. the central 
oxygen abundances in luminous galaxies with M$_{\rm B}$ $<$ --20.25 
do not 
show an appeciable correlation with galaxy luminosity. 
This flattening of the L -- Z relation 
can be seen more clearly in the bottom panel of Fig.~\ref{figure:lz} 
which shows a blow-up of the high-luminosity 
end of the relation. The presence of an upper envelope in the 
L -- Z relation at 
 12+log(O/H) $\sim$ 8.87, is clearly seen. 
The flattening of the L -- Z relation at high luminosities 
has been noted before by 
\citet{pilyuginetal04,tremontietal04}. 
 
What is the meaning of such a plateau? 
There are two known reasons for the existence of the L -- Z relation. 
First, it can be caused by a dependence of the efficiency of 
galactic winds to get rid of metals on the galaxy's luminosity:
more luminous and massive galaxies are less efficient in losing 
heavy elements by galactic winds. In this case, the L -- Z relation represents the ability of a 
given galaxy to retain the products of its own chemical evolution 
rather than its 
ability to produce metals \citep{larson74}. 
It is believed that the galactic winds do not play a 
significant role in the chemical evolution of the largest spiral galaxies
\citep[e.g.][]{garnett02,pilyuginetal04,tremontietal04}. 
The second reason that has been invoked to explain the L -- Z relation, 
which we adopt, is that the astration level increases and the gas mass 
fraction decreases as the luminosity of a galaxy increases 
\citep{mcgaugh97}. 
The existence of a plateau at high luminosities in the L -- Z relation thus 
implies that, on average, the central parts of large luminous 
spiral galaxies have 
similar astration levels.

Another prominent feature of the L -- Z relation is the 
rather large scatter 
in oxygen abundances at a given luminosity, $\Delta$log(O/H) $\sim$ 0.25. 
Part of this scatter may be due to uncertainties in 
the oxygen abundances. But another part is likely to be real. 
This scatter may be explained by fluctuations 
of the gas mass fraction $\mu$ among galaxies of a given luminosity. 
The simple model of chemical evolution of galaxies predicts that a 
decrease of $\mu$ by 0.1 results in an increase of the oxygen
abundance by $\sim$ 0.13 dex, in the range of $\mu$ from $\sim$ 0.50 to 
$\sim$ 0.05. Then, a scatter $\Delta$log(O/H) $\sim$ 0.25 may 
be explained by fluctuations of the gas mass fraction as large as 
$\Delta \mu$ $\sim$ 0.2 among galaxies of a given luminosity. 
The global gas mass fractions in the sample spiral galaxies have been 
estimated 
to be low, $\mu$ $<$ 0.25 \citep{garnett02,pilyuginetal04}. 
This suggests that the gas in the centers of the most metal-rich galaxies 
has been almost completely converted into stars. Consequently, the observed 
oxygen abundance in the centers of those galaxies 
represents the maximum attainable value of the 
oxygen abundance. This provides a natural explanation for the 
constant maximum value of the observed central oxygen abundance in  
the most oxygen-rich galaxies (Fig.~\ref{figure:lz}).

The maximum value of the gas-phase oxygen abundance in 
H\,{\sc ii} regions of spiral galaxies is thus 12+log(O/H) $\sim$ 8.87. 
Some fraction of the oxygen is locked into dust grains \citep{meyer98,estebanetal98}. 
According to \citet{estebanetal98}, the fraction of the dust-phase oxygen 
abundance in the Orion nebula is about 0.08 dex (but see \citet{simondiaz06}). 
Then, the maximum value of the gas+dust oxygen abundance in 
H\,{\sc ii} regions of spiral galaxies is 12+log(O/H) $\sim$ 8.95. 

\subsection{Discussion}

The oxygen abundances we have derived here for luminous spiral galaxies
are significantly lower than those obtained previously by a number 
of previous investigators  
\citep{garnett02,tremontietal04,melbourne02,lamareille04}.
This is not surprising.  
The abundances in the papers quoted above have been derived using   
different versions of the early calibration. 
It has been argued \citep{hcal,pilyuginetal04} 
that the oxygen abundances so determined are significantly 
overestimated at the high-metallicity end. Indeed, \citet{tremontietal04} 
have themselves noted that their oxygen abundances may have been
overestimated by as much as a factor of two.
On the other hand, the oxygen abundances obtained here 
are not based on any calibration since they 
are derived via the T$_{\rm e}$ method coupled with the ff relation. 
The ff relation is purely empirical in the sense that it relates  
directly measured quantities, without any other assumption. 

The central oxygen abundances of two luminous spiral galaxies, 
M~101 (M$_{\rm B}$ = --21.65) and M~51 (M$_{\rm B}$ = --21.48),
have recently been determined  
from measurements of (O/H)$_{\rm T_e}$ abundances in a number 
of H\,{\sc ii} regions. 
They have been found to be respectively 12+log(O/H)$_{\rm T_e}$ = 8.76 
\citep{kennicuttetal03}, and 12+log(O/H)$_{\rm T_e}$ = 8.72 
\citep{bresetal04}. 
Those data are in agreement with our L -- Z 
relation, but are in severe conflict with relations 
based on O/H abundances 
derived with the one-dimensional empirical calibrations. 

Comparison with H\,{\sc ii} region photoionization models has led some 
authors \citep[e.g.][]{stasinska05} to question the applicability of the 
classic T$_{\rm e}$ method to the high-metallicity regime. 
We have 
suggested \citep{vybor} using the interstellar oxygen abundance in the solar 
vicinity, derived with very high precision from high-resolution observations 
of the weak interstellar OI$\lambda$1356 absorption line towards the stars, as 
a "Rosetta stone" to check the reliability of the oxygen abundances derived 
in H\,{\sc ii} regions with the T$_{\rm e}$  method. The agreement between the 
value of the oxygen abundance at the solar galactocentric distance derived 
from the T$_{\rm e}$ method and that derived from the OI$\lambda$1356 
interstellar absorption line towards the stars provides strong support for 
the applicability of the classic T$_{\rm e}$ method to 
solar-metallicity objects. 
We have also examined previously 
the reliability of oxygen abundances derived in high-metallicity 
H\,{\sc ii} regions with the T$_{\rm e}$  method
by analyzing the radial 
distributions of oxygen abundances in the disks of spiral galaxies 
\citep{pilyuginetal05}. 
According to \citet{stasinska05}, the 
derived (O/H)$_{\rm T_e}$ value is very close to the real one 
as long as the metallicity is low. Discrepancies
appear for oxygen abundances 
12+log(O/H) $\sim$ 8.6 - 8.7, and may become very large as  
metallicity increases \citep[Fig. 1a in][]{stasinska05}. The derived 
(O/H)$_{\rm T_e}$ values are smaller than the real ones, 
sometimes by enormous factors. 
If this is the case, then 
the radial distribution of (O/H)$_{\rm T_e}$  abundances should 
show a bow-shaped curve with a maximum value of 12+log(O/H) $\sim$ 8.7 
at some galactocentric distance. However, the derived radial distributions of 
(O/H)$_{\rm T_e}$  abundances in the disks of spiral galaxies do not show 
such an  
appreciable curvature, and the (O/H)$_{\rm T_e}$  abundances 
increase more or less monotonically with decreasing galactocentric distance
\citep{pilyuginetal05}. 
Thus, while our results do not rule out the possible existence of the 
Stasi\'{n}ska's effect, they suggest that the great majority of H\,{\sc ii} 
regions in galaxies are in a metallicity range where this effect is not 
important. 
Another factor that may affect our abundance determinations is temperature 
fluctuations inside H\,{\sc ii} regions \citep{peimbert67}. If they are 
important, our abundances would be underestimated. 

The abundances derived here thus
depend on the ability of the classic T$_{\rm e}$ method to give correct 
abundances in the  
high-metallicity regime. They will 
be subject to revision if, when metal-rich H\,{\sc ii} regions are better 
understood, it is shown that the T$_{\rm e}$ method, because of temperature 
fluctuations, strong 
electron temperature gradients inside H\,{\sc ii} regions, or any other 
reason, 
results in incorrect abundances at high metallicities. 

The Sun is one of the widely used reference objects in astrophysics. 
Standard practice is to express the element content in a cosmic object 
via the corresponding value for the Sun, i.e. the composition of the 
Sun is used as standard unit. 
For many years, the recommended solar oxygen abundance was 
12+log(O/H)$_{\sun}$ $\approx$ 8.9. 
This high abundance was obtained from a one-dimensional hydrostatic model 
of the solar atmosphere.
Recently the solar oxygen abundance has been significantly 
reduced as a result of a time-dependent, three-dimensional 
hydrodynamical model of the solar atmosphere.
Taking the average of recent determinations
(12+log(O/H)$_{\sun}$ = 8.70 in \citet{allendeprietoetal04}, 8.66 in 
\citet{asplundetal04}, 8.59 in \citet{melendez04}), the solar abundance is now
12+log(O/H)$_{\sun}$ = 8.65. 
Thus, the maximum value of the gas+dust oxygen abundance of 
H\,{\sc ii} regions in spiral galaxies is higher by 
a factor of $\sim$ 2 than the solar value.

\section{The oxygen yield}

\subsection{Basic considerations}

The simple model of chemical evolution of galaxies predicts that the 
oxygen abundance of the interstellar matter of a galaxy is related to 
the gas mass fraction $\mu$ and the oxygen yield Y$_{\rm O}$ by
the following formula
\begin{equation}
Y_{\rm O} = \frac{Z_{\rm O}}{\ln (\frac{1}{\mu})}   .
\label{equation:y}
\end{equation}
In a real situation, the oxygen abundance is also affected by the mass exchange 
between a galaxy and its environment. This mass exchange can alter the above relation and 
mimic a variation in 
the oxygen yield. In that case, the simple chemical evolution model 
is used to estimate the ``effective'' oxygen yield $Y_{\rm eff}$ 
\citep{edmunds90,vila92}

As noted above, it is believed that galactic winds do not play a 
significant role in the chemical evolution of the largest spiral galaxies.
For these, is the oxygen yield derived by using 
Eq.~(\ref{equation:y}) close to the true oxygen yield? 
This may not be the case for two reasons. First, the simple model 
is based on the instantaneous recycling approximation. 
Oxygen is produced and ejected into the interstellar medium by 
massive stars with lifetimes much shorter than the evolution time of 
spiral galaxies. From this point of view, 
the instantaneous recycling approximation is justified. 
However, the value of Y$_{\rm O}$ 
depends not only on the amount of oxygen but also on the total mass of 
matter ejected in 
the interstellar medium (see below). In other words, the instantaneous 
recycling approximation assumes that the next 
generation of star is formed when all stars from previous generations 
have finished their evolution. This never occurs in a real galaxy.
As a consequence, the simple model predicts slightly higher oxygen abundances 
than numerical 
models for the chemical evolution of galaxies with realistic star formation 
histories \citep{azh}. Fortunately, this difference is small and can be 
neglected in the case of oxygen.
 
Furthermore, it is well known that the simple model predicts many more 
low-metallicity 
stars than are observed in the solar neighbourhood, the so called 
``G--dwarf'' paradox. Various versions of the infall model, in which an
infall of gas onto the disk takes place for a long time, 
have been proposed to account for the observed 
metallicity distribution in the solar neighborhood.
\citep[among many others]{tosi88a,tosi88b,pilyuginedmunds96a,pilyuginedmunds96b,chiappinietal01}. 
An infall model has also been applied to other spiral 
galaxies \citep[and references therein]{pagbook}.
It is thus generally accepted that gas infall plays an important 
role in the chemical evolution of disks of spiral galaxies. Therefore, the 
application of the simple model to large spiral galaxies to estimate the true 
oxygen yields may appear unjustified. It is expected that the 
rate of gas infall onto the disk decreases exponentially with time
\citep[e.g.][]{pilyuginedmunds96a,pilyuginedmunds96b}. 
It has been shown \citep{pilyuginferrini98} 
that the present-day location of a system in the $\mu$ -- O/H 
diagram is governed by its evolution in the recent past, but is 
independent of its evolution on long timescales. 
Therefore, the fact that the 
present-day location of spiral galaxies is near the one predicted by 
the simple model is not in conflict with a picture in which an infall of 
gas onto the disk takes place during 
a long time (the latter is necessary to satisfy the 
observed abundance distribution function and the age -- metallicity relation 
in the solar neighbourhood) since these observational data reflect the
evolution of the system in the distant past. 
Therefore, one can expect that the application of the simple model to large 
spiral galaxies, Eq.~(\ref{equation:y}), 
provides a more or less reasonable 
estimate of the true oxygen yield Y$_{\rm O}$. 
Certainly, to find an accurate 
value of the true oxygen yield Y$_{\rm O}$, an appropriate models of 
chemical evolution of galaxies should be computed.

\subsection{An empirical estimate of the oxygen yield}

From Eq.~(\ref{equation:y}), it is clear that an accurate determination of 
the oxygen yield depends on accurate oxygen abundance 
and gas mass fraction measurements. 
Our derived (O/H)$_{\rm ff}$ abundances are more accurate 
than the (O/H)$_{\rm R_{23}}$ abundances used in previous studies. 
However, a precise estimate of the gas mass fraction is not a trivial task,
and this for several reasons.
On the one hand, the mass of the stellar component of a galaxy is usually 
estimated by converting the measured luminosity to mass via an adopted 
mass-to-luminosity ratio. The latter is strongly 
model-dependent, therefore it is difficult to get a reliable estimate of 
it for individual galaxies. 
It is widely accepted that the converting factor from the 
near-infrared luminosity 
is a robust quantity for deriving the stellar mass of a galaxy. 
This quantity shows a low dependence on the star formation 
history but depends strongly on the initial mass function (for example, 
on the adopted lower stellar mass limit). 
On the other hand, 
the mass of molecular hydrogen can only be estimated by 
indirect methods. The commonly accepted method is the use of the CO line flux 
and a conversion factor $X$ between the flux in the CO line and the amount of 
molecular hydrogen. The conversion factor $X$ = N(H$_2$)/I(CO) 
depends strongly 
on the physical properties of the interstellar medium which are known to vary 
from galaxy to galaxy. 
The best-estimated values of X for 
a sample of well-studied nearby galaxies span the range 
from 0.6 to 10 $\times$ 10$^{20}$ mol cm$^{-2}$ (K km s$^{-1}$)$^{-1}$ 
\citep{bosellietal02}. 
Thus, the value of the oxygen yield derived from Eq.~(\ref{equation:y}) can be 
strongly affected by the uncertainty in the gas mass fraction determination.

\begin{figure}
\resizebox{1.00\hsize}{!}{\includegraphics[angle=000]{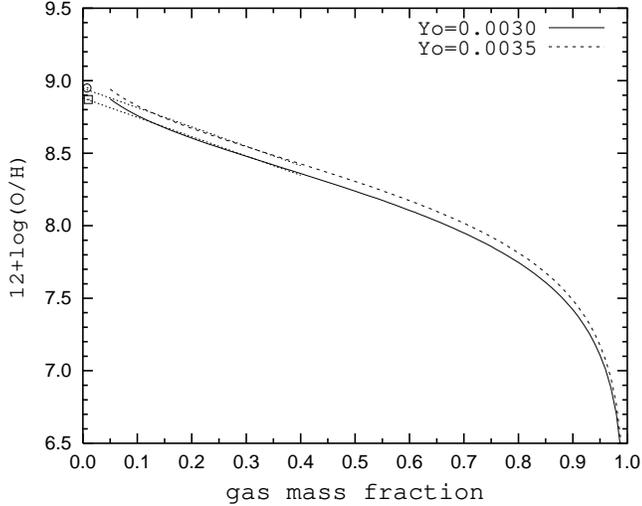}}
\caption{
The oxygen abundance as a function of the gas mass fraction predicted 
by the simple models of chemical evolution of galaxies, 
with an oxygen yield Y$_{\rm O}$=0.0030 (solid line) and with 
Y$_{\rm O}$=0.0035 (dashed line). The extrapolations to $\mu$=0 for these   
models are shown by the dotted lines. 
The open square is the maximum value of the observed gas-phase oxygen abundance in 
H\,{\sc ii} regions of spiral galaxies. 
The open circle is the maximum value of the gas+dust oxygen abundance.
}
\label{figure:cb}
\end{figure}

The use of the maximum value of the oxygen abundance derived here 
allows us to overcome 
the above problem in the following way. 
We compare the derived maximum value of the 
oxygen abundance in galaxies with oxygen abundances predicted by the simple 
models with different oxygen yields for $\mu$=0. 
Fig.~\ref{figure:cb} shows the oxygen abundance as a function of the 
gas mass fraction predicted by the simple model with Y$_{\rm O}$=0.0030
(solid line) and with Y$_{\rm O}$=0.0035 (dashed line). 
Since the oxygen abundances are expressed in units of number of oxygen atoms 
relative to hydrogen, 
while Z$_{\rm O}$ in Eq.~(\ref{equation:y}) have units of mass fraction, 
we adopt the following conversion equation for oxygen  \citep{garnett02}
\begin{equation}
Z_{\rm O} = 12 \frac{O}{H}  .
\end{equation}

The simple model breaks down as the gas mass fraction approaches zero because
the term ln(1/$\mu$) blows up. 
Therefore the oxygen abundance predicted by the 
simple model at $\mu$=0 should be estimated by extrapolation 
of the model predictions to $\mu$=0. The extrapolations of the simple  
models for two values of the oxygen yield are shown in 
Fig.~\ref{figure:cb} by the dotted lines. 
The fact that the central oxygen abundances in galaxies 
are derived here, not as oxygen abundances of H\,{\sc ii} regions in  
the very central parts of galaxies, but also 
as extrapolations of linear fits, 
justifies the above method.

The maximum value of the observed gas-phase oxygen abundance in 
H\,{\sc ii} regions of spiral galaxies is shown by the open square   
in Fig.~\ref{figure:cb}. The maximum value of the gas+dust oxygen 
abundance is shown by the open circle. 
Examination of Fig.~\ref{figure:cb} shows that the maximum value of the 
observed gas-phase oxygen abundance in H\,{\sc ii} regions of spiral galaxies 
corresponds to the simple model with Y$_{\rm O}$ $\sim$ 0.0030, and
the maximum value of the gas+dast oxygen abundance in spiral galaxies 
corresponds to the simple model with Y$_{\rm O}$ $\sim$ 0.0035. 

Thus, we can conclude that the value of the oxygen yield is about 0.0035,
 depending on the fraction of oxygen incorporated 
into dust grains. The value of the oxygen yield derived here from the gas-phase
oxygen abundance (Y$_{\rm O}$ $\approx$ 0.0030) is close to that obtained 
recently for spiral galaxies by \citet{pilyuginetal04} (Y$_{\rm O}$ $\approx$ 
0.0027) and by \citet{bresetal04} (Y$_{\rm O}$ $\approx$ 0.0032), but is 
significantly lower than the oxygen yield obtained by \citet{garnett02} 
(Y$_{\rm O}$ $\approx$ 0.010).

\subsection{The stellar oxygen yield}

\begin{figure}
\resizebox{1.00\hsize}{!}{\includegraphics[angle=000]{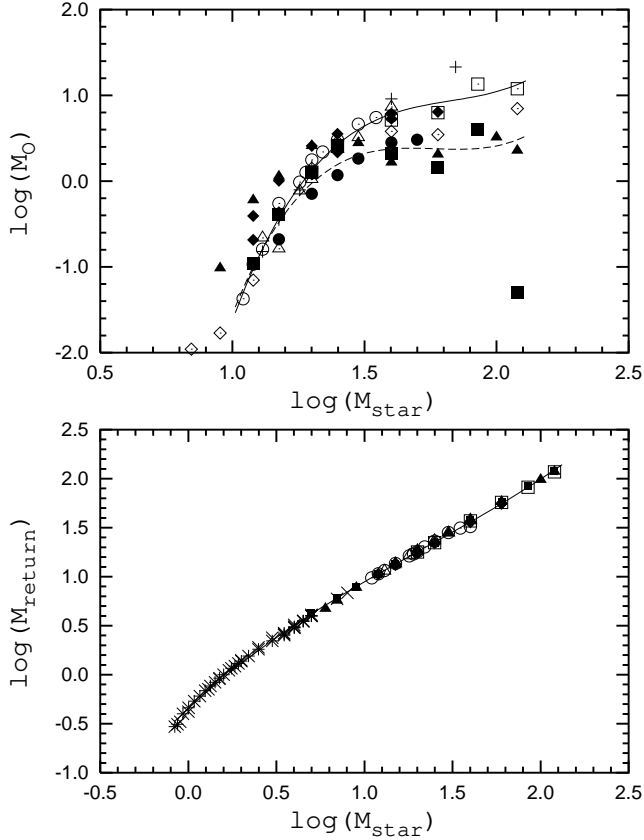}}
\caption{
{\it Top panel.} The mass of freshly manufactured oxygen ejected by a star 
in the interstellar medium as a function of the initial mass of the 
star. The results from \citet{maeder92} 
for models with low stellar wind mass loss rates 
are shown by open squares, and that for models with high stellar wind mass 
loss rates are shown by filled squares. 
Results from other investigators have also been plotted: 
\citet{lh95} (filled circles),  
\citet{nomotoetal97} (plus signs),  
\citet{portinarietal98} (filled triangles),  
\citet{meynetmaeder02} (open rhombus),  
\citet{hirschietal05} (filled rhombus),  
and \citet{kobayashietal06} (open triangles).
The solid line is a fit to the model results of 
\citet{maeder92} (models with low stellar wind mass loss rates),  
\citet{ww95}, \citet{nomotoetal97}, and \citet{kobayashietal06}. 
The dashed line is a fit to the model results of  
\citet{maeder92} (models with high stellar wind mass loss rates), 
\citet{lh95} and \citet{portinarietal98}. 
{\it Bottom panel.} The mass returned by a star to the interstellar medium 
as a function of the initial mass of the star.
The data for low- and intermediate-mass stars from \citet{vandenhoek97} 
are shown 
by crosses, those from \citet{marigo01} by open circles. The sources for  
the models and the symbols for massive stars are the same as 
in the top panel. 
The solid line is a fit to all the data.
}
\label{figure:om}
\end{figure}

Can the derived value of Y$_{\rm O}$ be reproduced by the existing stellar 
evolution models? The similarity between the present 
oxygen abundances in the interstellar 
matter in the solar vicinity and that in the Sun which was set up some 
4.5 Gyr ago, 
suggests that spiral galaxies have had a high oxygen abundance
during most of their evolution. We therefore
use estimates of the yields in the high-metallicity regime.  
We have compiled the yields at solar metallicity from several sources 
\citep{maeder92,ww95,lh95,nomotoetal97,portinarietal98,meynetmaeder02,hirschietal05,kobayashietal06}.
Following \citet{timmesetal95}, we have used models A from \citet{ww95} 
for stars in the 11 - 25 M$_{\odot}$ mass range, and models B 
for masses equal to 30, 35, and 40 M$_{\odot}$. The yields computed 
by \citet{meynetmaeder02} for the metallicity Z = 0.020 are given in 
\citet{cmm03}. 
The mass M$_{\rm O}$ of freshly produced oxygen ejected by a star in the 
interstellar medium, calculated by different authors, is shown  
as a function of the initial stellar mass 
in the top panel of Fig.~\ref{figure:om}. That figure shows that there exists
significant differences between the various predictions of M$_{\rm O}$
 at a given initial stellar mass. The differences are especially large for
stellar masses larger than $\sim$ 25 M$_{\odot}$. 
They are caused by differences in the input physics 
(stellar wind, mixing, etc) 
used by the various authors.
Close examination of the top panel of Fig.~\ref{figure:om} reveals that there 
are two main groups of points:  
a group of high M$_{\rm O}$ values from  
\citet{maeder92,ww95,nomotoetal97,kobayashietal06},
based on stellar models with small mass loss rates from stellar winds;  
and a group of low M$_{\rm O}$ values from 
\citet{maeder92,lh95,portinarietal98}, based 
on stellar models with large mass loss rates from stellar winds.
We will consider both the high and low values of M$_{\rm O}$.  
They can be considered as delimiting the range of oxygen production by stars.

It worth noting that the oxygen production is computed for 
single stars. A large fraction of stars are members of binary systems and 
evolve with mass exchange between components. Therefore one may expect that 
the stellar evolution models with strong stellar winds give 
values of the oxygen yield that are closer to the true value.

The cumulative mass $Q_{\rm O}$ of freshly manufactured oxygen ejected by a single stellar 
population at time $\tau$ after its formation is calculated as the sum of 
contributions from each star down to a stellar mass $M_{\rm S}^*$, 
corresponding to a lifetime $\tau$. 
It is given by the following expression 
\begin{eqnarray}
Q_{\rm O}(M_{\rm S}^*,M_{\rm U}) = \sum_{M_{\rm S}^*}^{M_{\rm U}}  
M_{\rm O} \, \varphi \, \Delta M_{\rm S}
\label{equation:qo}   
\end{eqnarray}
where $\varphi \, \Delta M_{\rm S}$ is the number of stars within the mass 
interval $\Delta M_{\rm S}$ as determined from the initial mass function
$\varphi$, and $M_{\rm O}$ is the ejected mass of freshly manufactured oxygen 
from a star with mass $M_{\rm S}$. 
Since stars with masses lower than 10 M$_{\odot}$ do not manufacture an 
appreciable amount of oxygen, the lower mass limit M$_{\rm S}^*$ 
in Eq.~(\ref{equation:qo}) has been set to 10 M$_{\odot}$. 
Difficulties arise in the computation of  
Q$_{\rm O}$  because of the limited number of stellar masses for which 
models exist. 
The gaps between the masses for which models have been calculated 
produce artificial jumps in the resulting 
Q$_{\rm O}$(M$_{\rm U}$,M$_{\rm S}^*$) \citep{pilyugin92,azh}. 
To overcome these difficulties, we have adopted the following approach. 
The dependence of the ejected mass of freshly 
manufactured oxygen M$_{\rm O}$ on stellar mass M$_{\rm S}$ 
is approximated by a polynomial of 
degree 3. 
We have adopted the approximation 
\begin{eqnarray}
       \begin{array}{lll}
\log {\rm M_O} & = & - 23.29 + 38.84 \, \log {\rm M_S}  \\
               &   & - 21.02 \, (\log {\rm M_S})^2 
                     +  3.84 \, (\log {\rm M_S})^3
       \end{array}
\label{equation:nowind}   
\end{eqnarray}
for the case of models with no or low stellar wind mass loss rates. 
This 
relation is shown by the solid line in the top panel of Fig.~\ref{figure:om}. 
As for the case of models with high stellar wind mass loss rates, 
we adopt the  approximation 
\begin{eqnarray}
       \begin{array}{lll}
\log {\rm M_O} & = & - 24.04 + 41.69 \, \log {\rm M_S}  \\
               &   & - 23.66 \, (\log {\rm M_S})^2   
                     +  4.46 \, (\log {\rm M_S})^3
       \end{array}
\label{equation:wind}   
\end{eqnarray}
 This 
relation is shown by the dashed line in the top panel of Fig.~\ref{figure:om}. 
Model 
points with large deviations were not used in the derivation of these 
two approximations. 

The dependence of the mass M$_{\rm ret}$ of the matter returned by a star to 
the interstellar medium on the initial mass of the star is shown 
in the bottom panel of Fig.~\ref{figure:om}. 
Inspection of the figure shows that that there is agreement between the 
M$_{\rm ret}$ values from different authors, so that  
all the data can be approximated by the following expression 
\begin{eqnarray}
       \begin{array}{lll}
\log {\rm M_{ret}} & = & - 0.344 + 1.666 \, \log {\rm M_S}  \\
                   &   & - 0.500 \, (\log {\rm M_S})^2
                         + 0.126 \, (\log {\rm M_S})^3
       \end{array}
\label{equation:mass}   
\end{eqnarray}
The cumulative mass $Q_{\rm ret}$ returned by a single stellar population is 
given by the following expression 
\begin{eqnarray}
Q_{\rm ret}(M_{\rm S}^*,M_{\rm U}) = \sum_{M_{\rm S}^*}^{M_{\rm U}}  
M_{\rm ret} \, \varphi \, \Delta M_{\rm S} .
\label{equation:qm}   
\end{eqnarray}

To compute $Q_{\rm O}$ and $Q_{\rm ret}$, we have to precise 
the initial mass function (IMF).  
We first consider the \citet{salpeter55} IMF
\begin{eqnarray}
\phi({\rm M_S}) =  {\rm c \, M_{S}^{-2.35}} .
\label{equation:salp}   
\end{eqnarray}
The coefficient $c$ is defined by normalizing the IMF over the whole 
mass interval
\begin{eqnarray}
\int\limits_{M_{L}}^{M_{U}} {\rm M_S} \, \phi({\rm M_S}) \, d{\rm M_S} = 1
\label{equation:normir}   
\end{eqnarray}
where M$_{\rm L}$ and M$_{\rm U}$ are the lower and upper mass limits. 
These mass limits are not known  
and are in fact free parameters. For example, \citet{romanoetal05}
adopt M$_{\rm U}$ = 100M$_{\odot}$ while \citet{kobayashietal06} choose 
M$_{\rm U}$ = 50M$_{\odot}$. 
Sometimes, the fraction $\zeta$ of the total stellar
mass locked up in stars with masses 
above 1M$_{\odot}$ is fixed, instead of M$_{\rm L}$. 
\citet{portinarietal98} have found that a ``good range'' for 
$\zeta$ is between 0.3 and 0.4. 

\begin{figure}
\resizebox{1.00\hsize}{!}{\includegraphics[angle=000]{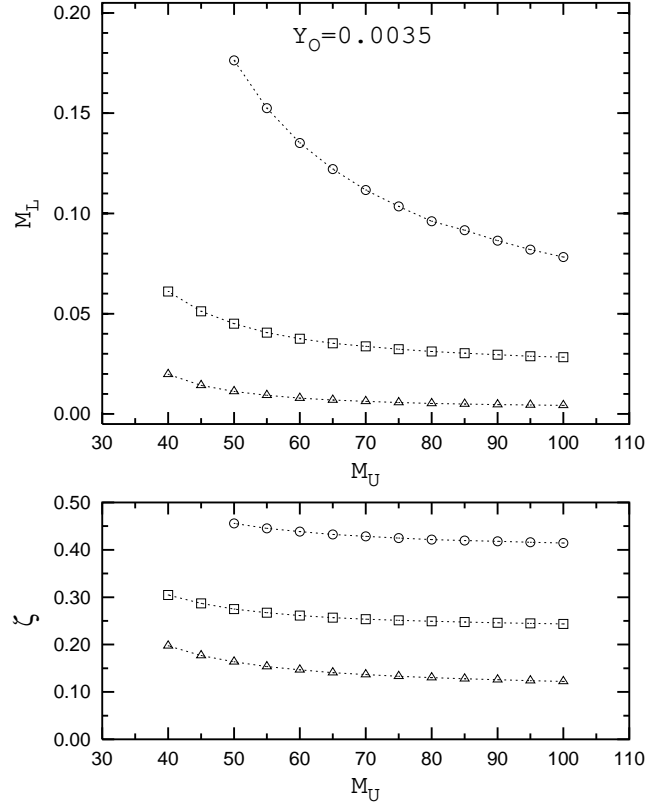}}
\caption{
{\it Top panel.} The lower mass limit M$_{\rm L}$ of the Initial Mass 
Function (IMF) 
calculated so that, for a given upper mass limit M$_{\rm U}$, the oxygen yield 
Y$_{\rm O}$ is equal to 0.0035. 
The circles correspond to Kroupa's IMF and oxygen yields given 
by stellar models with high stellar wind mass loss rates.
The squares correspond to a Salpeter IMF and oxygen yields given 
by stellar models with high stellar wind mass loss rates. 
The triangles correspond to a Salpeter IMF and oxygen yields given  
by stellar models with low stellar wind mass loss rates. 
Masses are in units of solar masses.
{\it Bottom panel.}  The fraction $\zeta$ of the total stellar mass 
in stars with masses  
above 1M$_{\odot}$ for the three combinations of 
IMF parameters and oxygen yields in the top panel. 
}
\label{figure:muml35}
\end{figure}

We next check whether there exist values of M$_{\rm L}$ and 
M$_{\rm U}$ so that  
Y$_{\rm O}^{\rm mod}$ = Y$_{\rm O}$, where the model oxygen yield 
Y$_{\rm O}^{\rm mod}$ is, by definition, the ratio of the mass of newly 
manufactured oxygen ejected by a single stellar population to the mass 
locked up in long-lived stars or remnants of that single stellar population 
\begin{equation}
Y_{\rm O}^{\rm mod} = \frac{Q_{\rm O}(M^*)}{1 - Q_{\rm ret}(M^*)} .
\label{equation:pi}   
\end{equation}
Since stars with masses 
less than 0.95 M$_{\odot}$ have lifetimes longer than the Hubble time 
\citep{schalleretal92}, we set M$^*$ = 0.95 M$_{\odot}$  
in Eqs.~(\ref{equation:qm},\ref{equation:pi}).
Then, by fixing Y$_{\rm O}^{\rm mod}$ = 0.0035 and M$_{\rm U}$, 
we can find  M$_{\rm L}$.
The resulting 
dependence of M$_{\rm L}$ on M$_{\rm U}$ is shown in the 
upper panel of Fig.\ref{figure:muml35}.
The triangles correspond to oxygen productions of 
stellar models with low stellar wind mass loss rates 
(Eq.~(\ref{equation:nowind})), while the squares correspond to 
oxygen productions of 
stellar models with high stellar wind mass loss rates
(Eq.~(\ref{equation:wind})). 
The lower panel of Fig.\ref{figure:muml35} shows by the same 
symbols the fraction $\zeta$ of the 
total stellar mass in stars with masses 
above 1M$_{\odot}$, for the same combinations of IMF parameters and 
oxygen productions  
as in the upper panel. 

A number of multi-component power-law stellar IMFs have been proposed as 
well. 
The one from \citet{kroupaetal93} is often used in the construction 
of chemical evolution models of galaxies. \citet{romanoetal05} have found 
that the Kroupa IMF fits better several observed 
properties of the solar vicinity as compared to the Salpeter IMF. 
The three-component power-law IMF of \citet{kroupaetal93} can be parameterized
as follows
\begin{eqnarray}
\phi({\rm M_S}) = \left\{
       \begin{array}{lll}
       {\rm c_1} \, {\rm M_S^{-1.3}} & {\rm for} & {\rm M_S} < 0.5 \, {\rm M}_{\odot} \\
       {\rm c_2} \, {\rm M_S^{-2.2}} & {\rm for} & 0.5 \, {\rm M}_{\odot} < {\rm M_S} < 1 \, {\rm M}_{\odot}  \\
       {\rm c_2} \, {\rm M_S^{-2.7}} & {\rm for} & {\rm M_S} > 1 \, {\rm M}_{\odot}  \\
       \end{array}
       \right.
\label{equation:kroupa}   
\end{eqnarray}
The coefficients $c_1$ and $c_2$ are defined by normalizing the IMF over the 
whole mass interval (Eq.~(\ref{equation:normir})), and by the condition 
$c_1\,{\rm M_S^{-1.3}} =  c_2 \, {\rm M_S^{-2.2}}$ for M$_{\rm S}$ = 
0.5M$_{\odot}$. 

The dependence of M$_{\rm L}$ on M$_{\rm U}$ for the Kroupa IMF 
and for oxygen productions from models with high stellar 
wind mass loss rates (Eq.~(\ref{equation:wind})) is shown  by circles in the 
top panel of Fig.\ref{figure:muml35}. 
The bottom panel shows also by circles the fraction $\zeta$ of the 
total stellar mass in stars with masses above 1M$_{\odot}$, for the same 
IMF parameters and oxygen productions as in the top panel. 
The value Y$_{\rm O}$ = 0.0035 cannot be reproduced with a Kroupa IMF 
and oxygen productions from models with low stellar 
wind mass loss rates. 

In summary, we have found that in the case of a Salpeter IMF, 
the observed yield, Y$_{\rm O}$ = 0.0035, 
can be satisfactorily reproduced by  
existing stellar evolution models with both low and high stellar wind mass 
loss rates. However, 
in the case of a Kroupa IMF, the observed yield can only be 
reproduced by stellar evolution models with high stellar wind mass loss 
rates. 
If the Kroupa IMF is more realistic than the Salpeter IMF, then our results 
suggest that stellar evolution models with low stellar wind mass loss rates 
predict too large oxygen yields. 
In the case of stellar models with high stellar wind mass loss rates, 
we obtain $\zeta$ 
$\sim$ 0.25 for a Salpeter IMF and $\zeta$ $\sim$ 0.40 for a 
Kroupa IMF. These values are close to the ``good range'',
 0.3 $<$ $\zeta$ $<$ 0.4, found by \citet{portinarietal98}. 

The above conclusion is at odds with 
the finding of \citet{crm03} that the oxygen yields in 
massive stars computed by 
\citet{ww95} without taking into account mass loss 
reproduce well abundance observations in the spiral galaxy M 101. 
The origin of the divergences in \citet{crm03} conclusions and ours 
probably comes from the fact that those authors used an early 
higher value of the central oxygen abundance of M 101, 
12+log(O/H) $\sim$ 9.2, while recent     
measurements of (O/H)$_{\rm T_e}$ abundances in a number of   
H\,{\sc ii} regions in M 101 have given a central 
oxygen abundance 12+log(O/H)$_{\rm T_e}$ = 8.76 \citep{kennicuttetal03}.

\section{Conclusions}

We search here 
for the maximum oxygen abundance in spiral galaxies. 
It is expected that this maximum value occurs at the 
centers of the most luminous galaxies. The luminosity -- central metallicity 
diagram for spiral galaxies is constructed. The central oxygen abundance in 
a galaxy is derived from a linear least-square best fit to the abundances 
in individual H\,{\sc ii} regions at various galactocentric distances. 
The oxygen abundance in an H\,{\sc ii} region 
is derived using the T$_{\rm e}$ method, 
with the flux in the [OIII]$\lambda$4363 
auroral line determined from the ff relation when not available from 
observations. The electron temperature t$_2$ 
is estimated from a newly derived t$_2$ -- t$_3$ relation \citep{tt}. 

We found that there exists a plateau in 
the luminosity -- central metallicity 
diagram at high luminosities (-22.3 $\la$ M$_B$ $\la$ -20.3). 
This provides strong evidence that the 
oxygen abundance in the centers of the most metal-rich luminous spiral 
galaxies reaches the maximum attainable value of oxygen abundance. 
The maximum value of the gas-phase oxygen abundance in H\,{\sc ii} regions of 
spiral galaxies is 12+log(O/H) $\sim$ 8.87. Because some fraction of the 
oxygen (about 0.08 dex) is expected to be locked into dust grains, the maximum 
value of the true gas+dust oxygen abundance in H\,{\sc ii} regions of spiral 
galaxies is 12+log(O/H) $\sim$ 8.95. This value is a factor of $\sim$ 2 higher 
than the recently estimated solar value. 

Based on our derived maximum value of the oxygen abundance in spiral 
galaxies, we have estimated the oxygen yield. We have 
found that the oxygen yield is around 0.0035,
depending on the fraction of oxygen incorporated into dust grains.  

\subsection*{Acknowledgments}

We thank the referee, Grazyna Stasi\'{n}ska, for helpful comments. 
This research was made possible in part by Award No. UP1-2551-KV-03 of the US 
Civilian Research \& Development Foundation for the Independent States of the 
Former Soviet Union (CRDF). 
   T.X.T. has been partially supported by NSF grant AST-02-05785. T.X.T. thanks 
the hospitality of the Institut d'Astrophysique in Paris and of the Service 
d'Astrophysique at Saclay during his sabbatical leave. He is grateful for a 
Sesquicentennial Fellowship from the University of Virginia.

\end{document}